\documentclass[11pt,twoside]{article}
\usepackage{asp2010}

\resetcounters

\newcommand{\lsim}{\mathrel{\hbox{\rlap{\lower.55ex \hbox{$\sim$}} \kern-.3em \raise.4ex \hbox{$<$}}}}

\newcommand{\kms}{km.s$^{\rm -1}$}
\newcommand{\bell}{$B_{\ell}$}
\newcommand{\bp}{$B_{\rm P}$}
\newcommand{\prot}{$P_{\rm rot}$}
\newcommand{\msun}{M$_{\odot}$}

\begin{document}

\title{The magnetic fields and magnetospheres of hot stars}
\author{Evelyne~Alecian,$^1$
\affil{$^1$LESIA-Observatoire de Paris, CNRS, UPMC Univ., Univ. Paris-Diderot, 5 place Jules Janssen, F-92195 Meudon Principal Cedex, France}}

\begin{abstract}
Strong advances in direct evidence of magnetic fields in hot massive stars have been possible thanks to the new generation of high-resolution spectropolarimeters such as ESPaDOnS (on the Canada-France-Hawaii Telescope) or HARPSpol (on the 3.6m ESO telescope). UV and optical high-resolution spectroscopy has also been very useful to study the magnetospheres of massive stars. In this contribution I review the observing tools and our current knowledge concerning the detection and characterisation of the magnetic fields and magnetospheres in hot stars.
\end{abstract}

\section{Introduction}

\subsection{Magnetic fields measurements and modelling}

The magnetic fields of stars can be detected and measured via the Zeeman effect. In presence of a magnetic field, the energy levels of an atom split into multiple equally-spaced sub-levels. In the most common case (see e.g. \citet{landi92} for more details), instead of a single line produced into a spectrum at the wavelength $\lambda_0$, a Zeeman triplet is formed. The triplet is composed of a linearly polarised component $\pi$ at the wavelength $\lambda_0$ and two right and left circularly polarised $\sigma_{\rm r,b}$ components at the wavelengths $\lambda \pm \bar{g}\lambda_{\rm B}$, where $\bar{g}$ is the effective Land\'e factor, and $\lambda_{\rm B}$ is the Zeeman splitting defined by:
\begin{equation}
\lambda_{\rm B}=\frac{e}{4\pi mc^2}\lambda_0^2B
\end{equation}
where $e$ is the electron elementary charge, $m$ is the electron mass, $c$ is the velocity of light, and $B$ is the magnetic strength.

In the peculiar cases of stars with very low projected rotational velocities ($v\sin i\lsim 5$~\kms, i.e. very narrow spectral lines) and strong magnetic fields (typically larger than $\sim$10~kG), we can either detect the three components of the Zeeman triplet or a widening of the line. However, in most cases, the Zeeman effect is not measurable inside the intensity lines due to the rotational Doppler broadening. However we can use the polarised properties of the Zeeman effect by recording the Stokes $V$ (circular polarisation), $Q$ and $U$ (linear polarisation) spectra. In weak ($B<30$~kG) magnetised stellar atmospheres, the circular polarisation is predicted to be one order of magnitude larger than the linear polarisation. Therefore, in most cases the circular polarisation only is measurable. However in few highly magnetised stars we have been able to measure the linear polarisation inside spectral lines. 

The level of circular polarisation is in most cases very low (about 0.01\% of the continuum intensity of a star), which makes it very difficult to detect. Furthermore the instrumentation has to be be developed in a way to avoid instrumental polarisation. Magnetic fields have been measured in many stars since the first measurement by \citet{babcock47}. Most of them were found in the chemically peculiar Ap/Bp stars and were typically stronger than $\sim1$~kG. The most common old technique was to measure the circular polarisation inside the wings of the Balmer lines using low-resolution spectropolarimeters. However \citet{bagnulo12} have recently shown that low-resolution spectropolarimetry can be very inaccurate when measuring relatively low magnetic fields. Only the high-resolution spectropolarimetry technique is able to reliably detect and provide accurate measurements of magnetic fields in stars of many kind and of various magnetic strengths and topologies. Today, the three most efficient high-resolution spectropolarimeters are ESPaDOnS on the 3.6m Canada-France-Hawaii Telescope (CFHT, Hawaii), Narval on the Telescope Bernard Lyot (TBL, Pic du Midi, France) and HARPSpol on the 3.6m-ESO Telescope (La Silla, Chili). Thanks to large telescope apertures, to improved efficiencies of polarimetric modules and spectrographs, and to multi-line analysis techniques, such as the Least-Squares Deconvolution \citep[LSD, ][]{donati97} technique, these instruments allow us to measure magnetic fields as low as few gausses in bright A-type stars \citep{petitp11}, and typically about 100 G in many kind of stars brighter than a magnitude 12 \citep{donati09}.

\begin{figure}
\includegraphics[angle=90,scale=0.47]{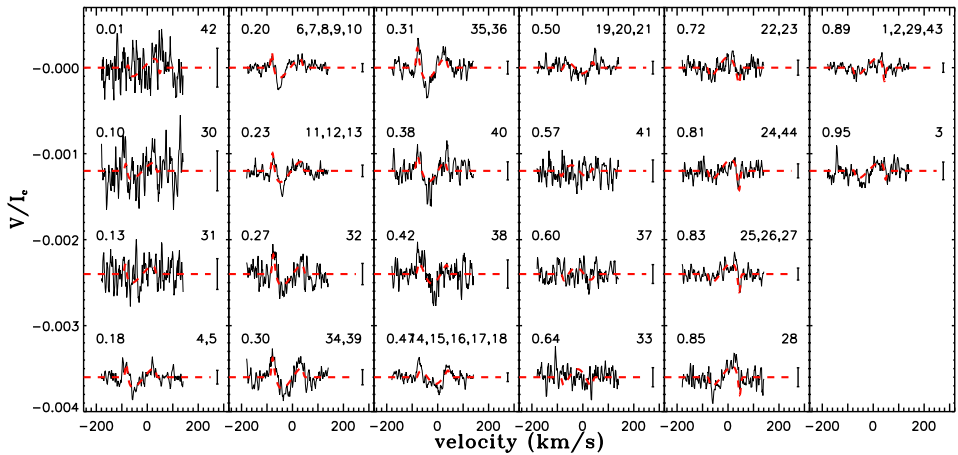}
\caption{Off-centred dipole best model (red dashed lines) superimposed on phase-binned LSD $V$ profiles of the He-strong star V2052 Oph. The rotation period was previously derived from UV data \citep{neiner03}. The best model parameters are: $i=53\pm8$\deg, $\beta=52\pm6$\deg, $B_{\rm P}=-418\pm46$~G, $d_{\rm dip}=0.18\pm0.04$~R$_*$ \citep[adapted from][]{neiner12}.}
\label{fig:fitv}
\end{figure}

In the weak field approximation, the Stokes $V$ parameter is proportional to the longitudinal magnetic field ($B_{\ell}$), i.e. the stellar averaged line-of-sight component of the magnetic field \citep{landi92}:
\begin{equation}
V = -C\bar{g}\lambda_{\rm 0}^2B_{\ell}\frac{dI}{d\lambda}
\end{equation}
where $C = e/(4\pi mc^2)=4.67\times10^{-13}\AA^{-1}$G$^{-1}$, and $dI/d\lambda$ is the wavelength derivative of the intensity inside the spectral line. In the case of an oblique rotator, i.e. a star rotating around an axis with an inclination $i$ from the line-of-sight, and hosting a dipole of strength \bp\ at the pole, with an obliquity $\beta$ from the rotation axis, the strength of \bell\ is changing as the star rotates. We can show that the curve of \bell\ describes a sinusoid \citep{landstreet70}. We therefore need to observe the star at different rotation phase in order to retrieve the magnetic configuration of the star. If the inclination of the system is known, by fitting the variations of \bell\ with a sinusoid we can derive the magnetic strength \bp\ and the obliquity $\beta$ \citep{borra80}. However, in most cases the inclination of the star is unknown. We can therefore take advantage of the resolution of the stellar surface provided by the Doppler broadening of spectral lines, and observe the variations of the Stokes $V$ signature within each pixel. In the case of an oblique rotator, with a centred dipole, for a set of parameter $(P_{\rm rot},i,\beta,B_{\rm P})$, \prot\ being the rotation period of the star, the Stokes $V$ signatures within the spectral line varies in a unique way. In many cases a centred dipole cannot properly fit the data, and an off-centred dipole, with respect to the centre of the star is required. In those cases an additional parameter $d_{\rm dip}$, i.e. the distance between the centre of the star and the dipole, expressed in stellar radius, is added to the model. Thanks to today's instrumentation we are now able to detect the Zeeman $V$ signatures inside the spectral lines of many intermediate and high-mass stars, record their variations at different rotation phase and retrieve the properties of the dipole that they host (see an example in Fig. \ref{fig:fitv}).

This method has been successful for many star hosting fields of simple configuration. However in some cases the magnetic configuration at the surface of the star cannot be approximate with a dipole. In those cases we can perform a detailed Zeeman (or Magnetic) Doppler Imaging (ZDI or MDI) by using the information that provides the Stokes $V$ spectra but also the linearly polarised Stokes $Q$ and $U$ spectra, which are sensitive to the transverse components of the magnetic field. By monitoring the star at various rotation phases in the three Stokes parameters we get the information on the three components of the magnetic field. By using the technique of magnetic inversion we can find a unique solution, and get a 2D map of the magnetic vectors and strengths at the surface of the star \citep[e.g.][]{kochukhov10}.

\subsection{Context: the magnetic field of low- and intermediate-mass stars}

The solar-mass stars possess a convective envelope surrounding a radiative core. Due to an interplay between the differential rotation and the convective motion inside the envelope, a dynamo field of solar-type is produced and localised inside stellar spots. The resulting surface magnetic fields are of complex configuration, highly variable on timescales of months, strongly correlated with stellar properties (such as the rotation period or age) and ubiquitous among low-mass stars ($M<1.5$~\msun) \citep{donati09}.

On the contrary, the intermediate-mass stars ($1.5<M<8$~\msun) are mainly radiative with a small convective core, and yet, some of them are magnetic. However their magnetic properties are very different from the solar-type stars: they are found only in the chemically peculiar Ap/Bp stars, they are organised on large scales (mainly dipolar or quadrupolar), they are strong (between 300 G and 30 kG), stable over many years, not correlated with any stellar property, and rare (between 1 and 10\%) \citep{donati09}. These properties are very different from the dynamo fields ones. This suggests that the origin of magnetic fields in intermediate-mass stars is very different from that of low-mass stars. The favoured hypothesis, today, is the fossil origin, i.e. the magnetic fields of the Ap/Bp stars are assumed to be either remnants of Galactic fields that have been swept during the molecular contraction phase, or remnants fields generated during stellar formation \citep{moss01,braithwaite06}.

According to the fossil field hypothesis a small fraction of the pre-main sequence intermediate-mass stars, i.e. the Herbig Ae/Be (HAeBe) stars, should host strong magnetic fields of simple configuration. In order to test this theory we performed a high-resolution spectropolarimetric survey of 128 HAeBe stars located in the field of the Galaxy and in three young clusters (NGC 6611, NGC 2244, NGC 2264). We used the ESPaDOnS and Narval instruments to record about 200 polarised spectra (one or two per target). We found 9 magnetic stars \citep{alecian12}, implying an incidence of about 10\%. Four of them have been studied in details, and we find that the fields are mainly dipolar, strong (from 300 G to  2.1 kG) and stable over more than 5 years \citep[e.g.][]{folsom08,alecian09}. These results confirm that a fossil link exist between the pre-main sequence and main-sequence magnetic fields in the intermediate-mass stars, and that their magnetic fields must have been shaped during the star formation.

\section{Magnetism in Massive Stars}

\subsection{Problematics}

The massive stars ($M>8$~\msun) have internal structures similar to the intermediate mass stars, i.e. a convective core and a large radiative envelope. One of the question that we can raise is: do they also host fossil magnetic fields ? In 2007, our knowledge of the magnetic properties of the massive OB stars ($M> 8$~\msun) was limited to isolated cases. Magnetic fields were detected mainly in chemically peculiar Bp (He-weak or He-strong) stars, but also in few magnetic massive stars with little or no chemical peculiarities \citep{donati09}. A couple of known magnetic stars displayed H$\alpha$ variable emission, variable photometry with eclipses, variable UV wind lines (e.g. C IV or N V), all with a period similar to the stellar rotation period. One of them is $\sigma$ Ori E \citep{landstreet78}, a B2V He-strong star of about 9~\msun, showing eclipses in its light-curve . The behaviour of the circumstellar H$\alpha$ variations of $\sigma$ Ori E, as well as a variable shell spectrum appearing only during the eclipses, lead to the proposition that a circumstellar plasma is confined into clouds in the close environment of the star, and is forced to corotate with the star by the magnetic field (e.g. Shore 1993). Following this picture \citet{townsend05} proposed a Rigidly Rotating Magnetosphere (RRM) model to explain the observed properties of the circumstellar emission of hot magnetic OB stars. This model assumes that streams of material in the stellar radiatively driven wind are leaving the surface at two opposite foot-points, are channelled along the magnetic lines, collide at the top of closed magnetic loops to form emitting hot plasma clouds, at the intersection of the rotation and magnetic equators. The material in clouds either falls back onto the star, or, if the star is rotating fast enough, stays confined and accumulates over the time. \citet*{townsend05b} have shown that the RRM model can very well reproduce the H$\alpha$ variations of $\sigma$ Ori E (see also the contribution of R. Townsend in this book).

Many questions have been raised in 2007 on the magnetic fields of massive stars. First, while we highly suspect that the origin of the magnetic fields is fossil, it still needs to be proven. While the magnetic fields observed in few OB stars look similar to those of Ap/Bp stars, no global picture of the magnetic properties of massive stars were available to test this hypothesis. Magnetic fields were observed in many Bp stars, but also in some non-peculiar stars, and the role of magnetic fields in building chemical peculiarities has to be understood. Finally we were wondering if the magnetosphere of $\sigma$ Ori E as proposed by Townsend et al. was a peculiar case or if it was the prototype of many if not all magnetic OB stars.

\subsection{The MiMeS project}
In order to investigate these questions, we started a large international project called Magnetic in Massive Stars (MiMeS\footnote{http://www.physics.queensu.ca/$\sim$wade/mimes}). The project is leaded by G. Wade in Canada and C. Neiner in France. The main observational part of this project is the compilation of a large number of high-resolution spectropolarimetric observations that we are acquiring since 2008 with ESPaDOnS, Narval and HARPSpol, thanks to three Large Programmes that G. Wade, C. Neiner, and myself have obtained, respectively. The observing sample has been divided into two sub-samples: the survey component (SC) that is intended to search for new magnetic stars among a sample of about 400 OB stars, and the targeted component (TC) that contains stars that were previously known to be magnetic, or that were part of the SC sample and have been discovered to be magnetic. We intend to acquire a large number of spectropolarimetric observations well sampled over the rotation phase of the TC stars, in order to study into details their magnetic fields and circumstellar environment. The data acquisition will end at the beginning of 2013. I will summarise the main results that the MiMeS collaboration obtained so far. I also recommend the reader to consult G. Wade's contribution of this book.

\subsection{MiMeS results: Magnetic detections and associated spectral peculiarities}
Within the MiMeS project, we have confirmed the magnetic files of 6 highly suspected magnetic OB stars, and detected 14 new magnetic stars among a sample of 320 stars, including six O-type stars which triple the number of known magnetic O-type stars \citep{grunhut09,alecian11,wade12}. One of the very interesting results of the MiMeS collaboration is the systematic detection of a magnetic field among the Of?p stars. Up to now, only five Galactic Of?p stars are known. All of them have been observed using high-resolution spectropolarimeters before and within the MiMeS project and all of them have been detected as magnetic \citep[][Grunhut et al. in prep., Shultz et al. in prep.]{donati06,martins10,wade12}. The main characteristics of the Of?p stars are recurrent spectral variations in the Balmer, He~\textsc{I}, C~{\sc III} and Si~{\sc III} lines, as well as strong emissions in the C~{\sc}III~$\lambda$~4650~\AA\ line \citep[see][for more details on Of?p stars]{naze08,naze10}. If this systematic detection is confirmed with a larger sample it would validate that the very specific spectral classification Of?p is an indirect indicator of magnetic fields. It is not yet clear how magnetic fields can explain all the spectral characteristics of that type of stars. The periodic variations, and especially those observed in the Balmer lines, can be explained as in $\sigma$~Ori~E by a magnetic oblique rotator with a co-rotating magnetosphere \citep[e.g.][]{martins10}.

MiMeS observations have investigated the case of hot B stars showing periodic modulations in wind-sensitive UV lines such as the Si~{\sc iv}~$\lambda$~1400~\AA\ and C~{\sc iv}~$\lambda$~1550~\AA\ lines \citep[see e.g. Fig. 2][]{neiner03}. It was proposed that surface magnetic fields could be at the origin of these variations \citep[e.g.][]{henrichs03}. V2052 Oph is one of the prototype of that kind of stars. \citet{neiner03} report the detection of a dipolar field at the surface of this star, that we have recently modelled within the MiMeS collaboration \citep[][and Fig. \ref{fig:fitv}]{neiner12}. Following those works, we have selected some OB stars showing strong variations in the C~{\sc iv} line and observed them with ESPaDOnS, Narval or HARPSpol. These observations lead to the discovery of two new magnetic stars stars: HD 25558 and $\sigma$~Lupi (Henrichs et al. in prep.). The magnetic analysis of these stars allowed us to derive the rotation period, and we confirm that the variations of the C~{\sc iv} line have the same period as the rotation period of the star. These results seem to confirm the presence of a confined wind, forced to corotate with the star by the stellar magnetic field.

\subsection{The dynamical versus centrifugal co-rotating magnetospheres of OB stars}

\begin{figure}
\includegraphics[scale=0.195]{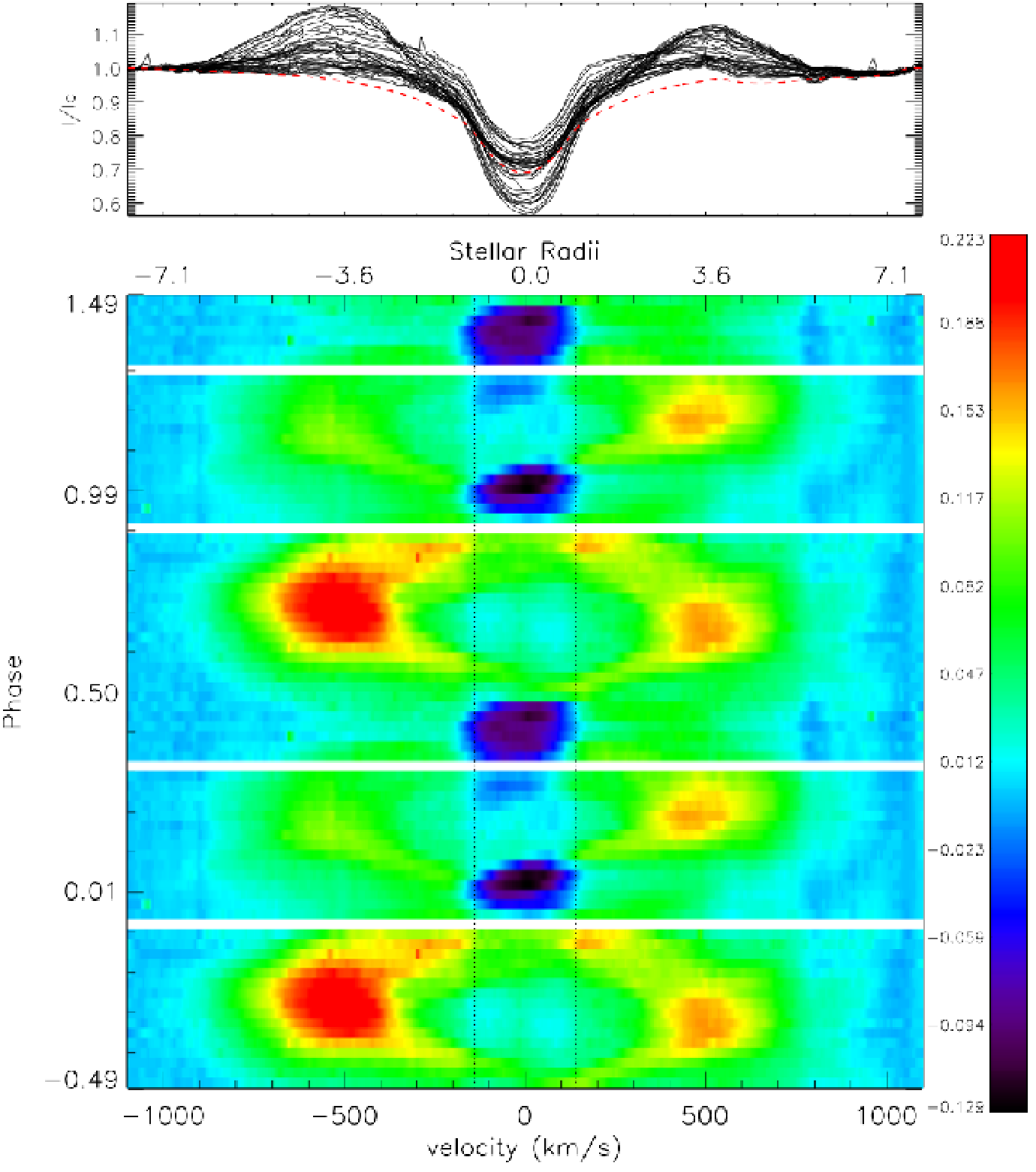}
\includegraphics[scale=0.19]{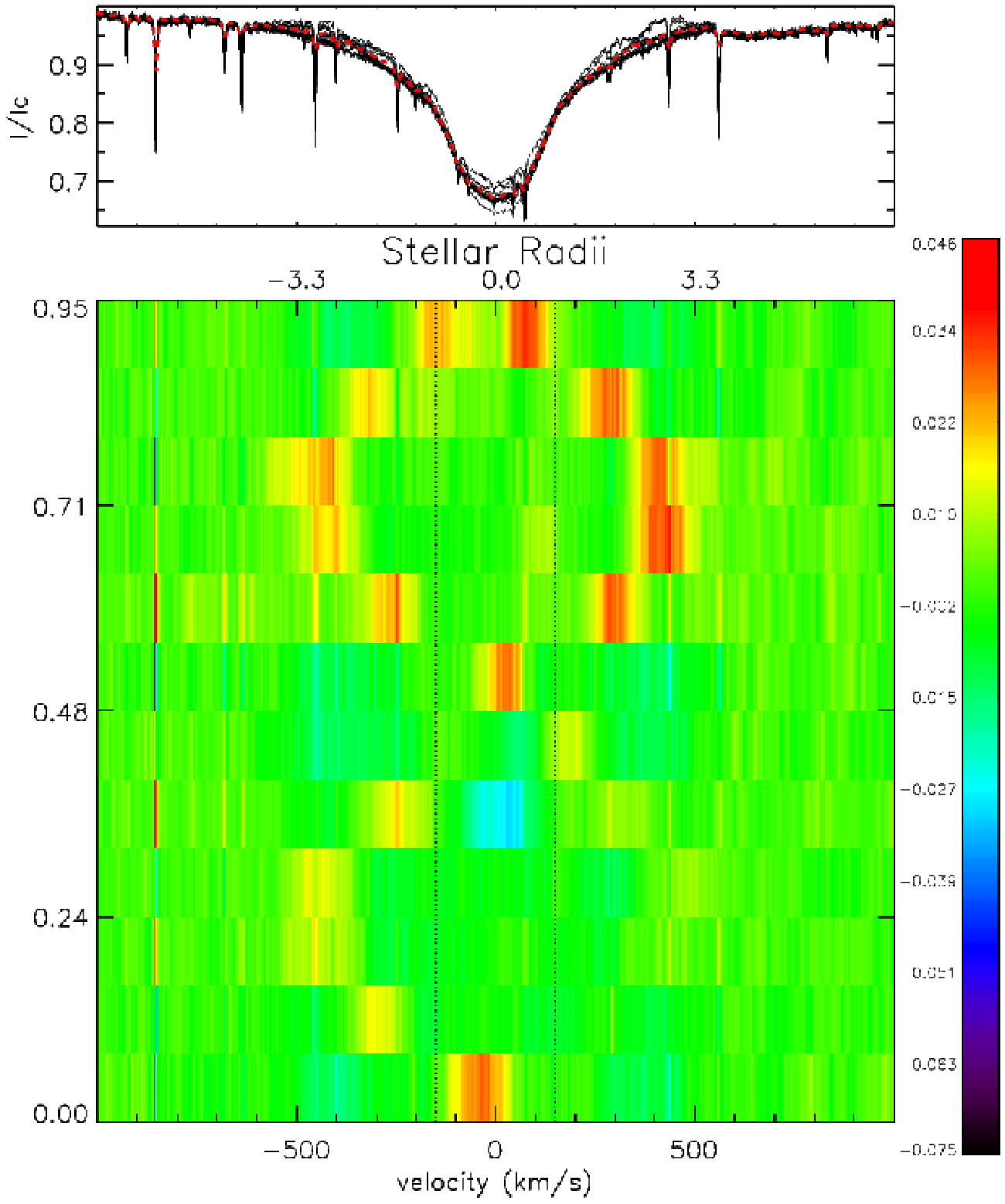}
\includegraphics[scale=0.20]{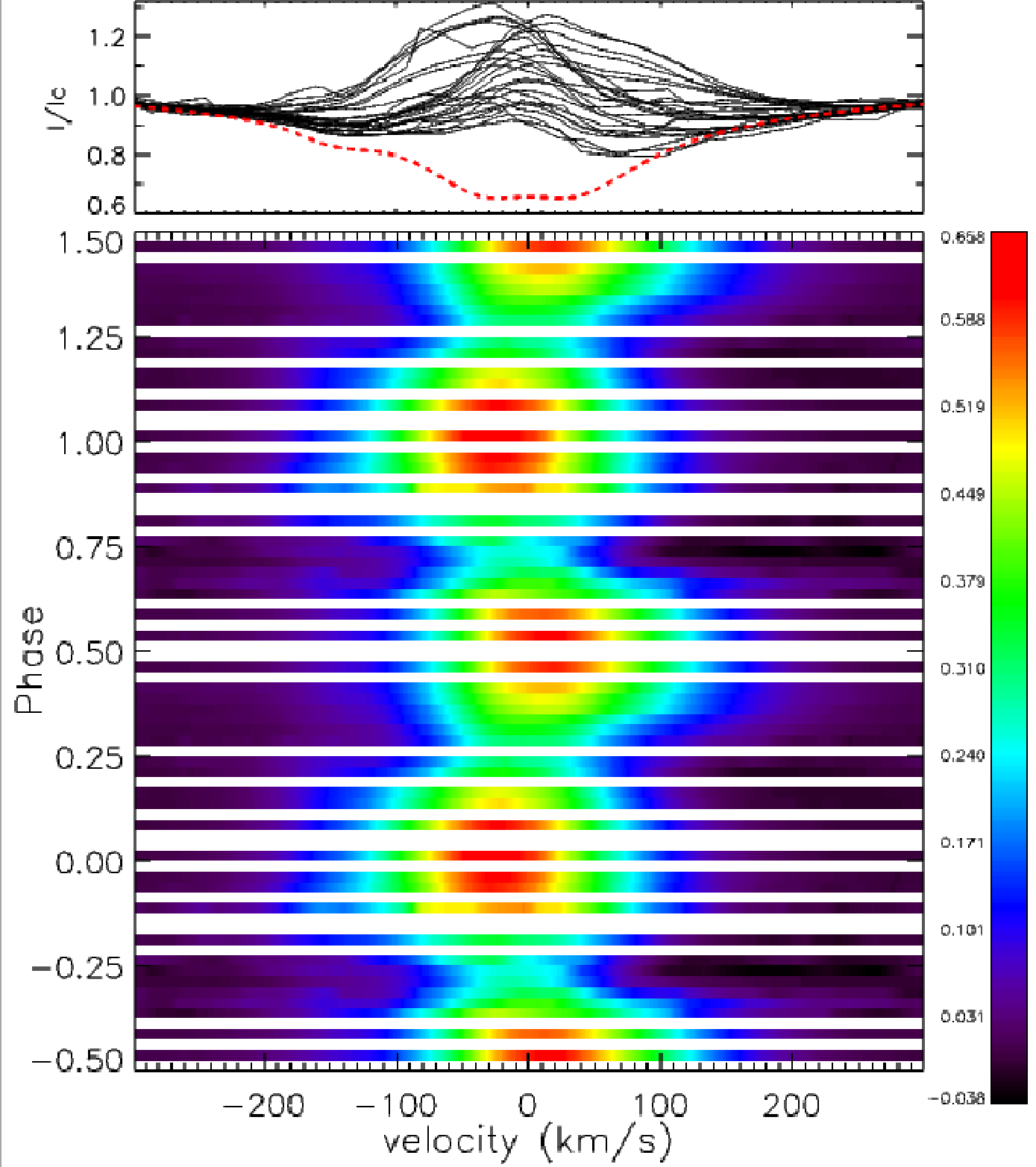}
\caption{Dynamic spectral plots of the B2 V star $\sigma$~Ori~E \citep[left,][]{oksala12}, the B3 V star HD 64740 (middle, Peralta et al. in prep.) and the O9 V star HD 57682 (right, Grunhut et al. submitted). In the case of $\sigma$ Ori E and HD 57682 the spectra are plotted minus a synthetic photospheric spectrum, while the spectra of HD 64740 have been plotted minus the mean of all observations.}
\label{fig:ha}
\end{figure}

In many magnetic B- and O-type stars, emissions have been detected in H$\alpha$, varying periodically with the same period as the rotation of the star. It therefore seems that co-rotating magnetospheres, similar to $\sigma$ Ori E, are present around many other magnetic OB stars (Fig. \ref{fig:ha}. Thanks to the MiMeS observations that allowed to compile a large quantity of spectra well sampled over the rotation phase for many magnetic stars, we find that the variations of the H$\alpha$ emission display a variety of behaviour that are not all similar to $\sigma$ Ori E. This can be explained by differences from star to star in magnetic obliquity and rotation axis inclination, as well as in rotation velocities that control the Doppler resolution of the CS environment in H$\alpha$.

We also find that H$\alpha$ emission is not ubiquitous among magnetic OB stars. To understand this, we have to consider the energetic balance that drives the formation and dynamic of rigidly-rotating magnetospheres. I will only present here, in a simple way, the physic of a $\{{\rm star}+{\rm magnetosphere}\}$ system, and will remind the reader the two main parameters that are of interest here. For more details I refer the reader to ud-Doula's and Owocki's contributions of the same book, as well as \citet*{uddoula02} and \citet*{uddoula08}. The presence or not of a co-rotating magnetosphere around a magnetic hot star is the result of the interplay between line-driven wind, rotation and magnetic field. The $\eta_*$ parameter ($\eta_*=B_{\rm eq}^2R_*^2/\dot{M}v_{\inf}$) is the ratio of the wind kinetic energy to the field energy at the surface of the star, and controls the magnetic confinement of the wind. In order to channel the matter down to the magnetic equators the magnetic energy has to dominate over the wind energy, and therefore $\eta_*$ has to be larger than one. If the star rotates slowly and the rotational energy can be ignored, once the matter has reached the magnetic equator at few stellar radii from the stellar surface, it can either leaves the system if it reaches a radius larger than the Alfven radius ($R_{\rm A}\sim \eta^{1/4}_*R_*$), or falls back onto the star due to the gravitational attraction of the star. The result is the formation of a dynamical magnetosphere (DM) between the stellar surface and the Alfv\'en radius, constituted of unstable disks or clouds, depending of the magnetic obliquity. The matter inside the magnetosphere are constantly falling back onto the star, and are replenished by new material arriving from the poles. However, if the system is rotating fast enough, the centrifugal force can prevent the matter from falling back onto the star. This can happen if the centrifugal force is stronger than the gravity, i.e. at radii larger than the Kepler radius as defined by ud-Doula et al. (2008): $R_{\rm K}=W^{-2/3}R_*$, where $W=v_{\rm rot}/\sqrt{GM_*/R_*}$ is the ratio of the equatorial surface rotation velocity over the orbital velocity near the equatorial surface. If the matter from the wind reaches the magnetic equator at radii between $R_{\rm K}$ and $R_{\rm A}$, it can pile up and form a stable structure that we call a centrifugal magnetosphere (CM). The RRM model therefore predicts that magnetic stars with $\eta_* < 1$ cannot develop co-rotating magnetospheres, stars with $W < \eta_*^{-3/8}$ can develop co-rotating dynamical magnetospheres and stars with $W > \eta_*^{-3/8}$ can develop co-rotating centrifugal magnetospheres.

\subsection{Confronting the RRM models with the MiMeS results}

In order to test these predictions, we have to find a way to differentiate the CM from the DM. In the case of a CM, the matter piles up and can therefore form clouds sufficiently dense to be easily detected in H$\alpha$, while in the case of a DM, the structures are highly dynamical and the material cannot pile up. A DM is therefore more difficult to detect in H$\alpha$. However the wind density is strongly dependent of the temperature of the star, and therefore we expect a DM magnetosphere to be more easily detected at high stellar temperature.

\begin{figure}
\includegraphics[scale=0.50]{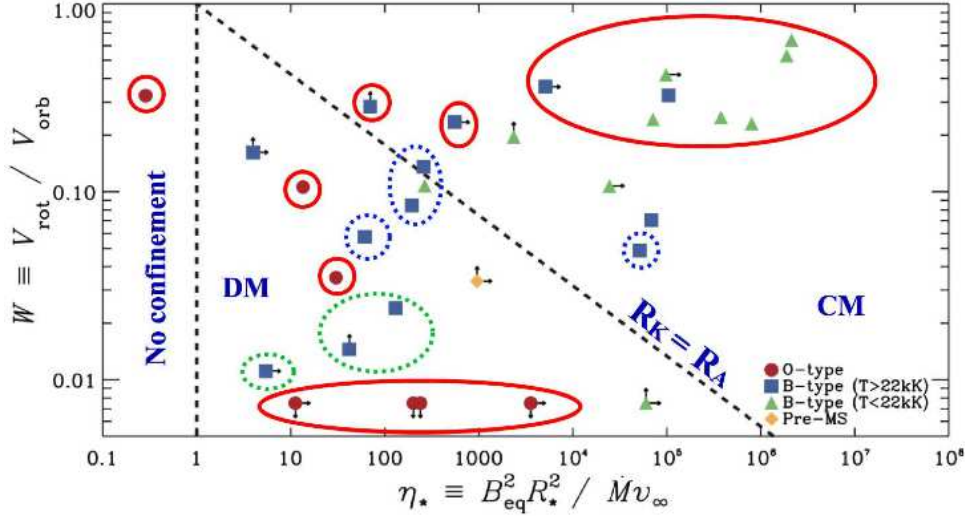}
\caption{The MiMeS magnetic OB stars plotted in the confinement-rotation diagram. The type of the symbols are dependent of the temperature of the stars. The stars surrounded by a red full line display variable H$\alpha$ emission. The blue and green dashed lines suround stars showing evidence of confined winds, but no H$\alpha$ emission. (Petit \& Owocki et al. in prep.)}
\label{fig:ipod}
\end{figure}

These predictions can be tested by separating the magnetic OB stars in the above categories (no confinement, DM and CM), and look for the signature of magnetospheres in their H$\alpha$ profiles. This is done in the confinement-rotation diagram displayed in Fig.~\ref{fig:ipod}. The vertical line separates two regions of the diagram where stars can confine or not their winds. The oblique line separate two regions where stars can develop dynamical or centrifugal co-rotating magnetospheres. The red dots represent the magnetic O-type stars, and the blue and green symbols represent the magnetic B-type stars. The red circles surround stars displaying H$\alpha$ variations. We first observe that all O-type stars are situated in the DM part of the diagram, which is due to their relatively slow rotation periods and faint magnetic fields compared to B-type stars. Secondly we remark that almost all of the stars situated in the CM region display variable H$\alpha$ emission as expected. The few ones that seem to not show H$\alpha$ emission are either cooler or are slower rotators, which makes the detection of CS matter in H$\alpha$ more difficult. In the DM part of the diagram we observe that all O-type stars display variable H$\alpha$ emission while none is detected in B-type stars, confirming the prediction of a correlation between H$\alpha$ emission and stellar temperature in the case of dynamical magnetospheres.

H$\alpha$ is not the only indicator of the presence of confined material around a star. The dashed blue and green lines surround stars displaying cyclical UV wind line variability, i.e. stars with evidence of a magnetically controlled wind. The wind of these B-type stars might not be dense enough to develop detectable DM or CM magnetospheres in H$\alpha$, but still seem to be confined by the magnetic field, and therefore consistent with the predictions of the RRM model. Finally for the symbols that are not surrounded by any kind of lines, it means that no CS emission is detected in H$\alpha$, but information on the UV wind lines are missing, and we are not able to check the confinement of their wind.

\section{Summary}

Thanks to the MiMeS project, our knowledge on the magnetic fields of massive OB stars have improved considerably the last few years. Similarly to the main-sequence and pre-MS intermediate mass stars, about 10\% of the OB stars host strong and organised on large scales (dipole or quadrupole) magnetic fields, which are very likely of fossil origin. Similarly to the A/B stars, the magnetic OB stars are found among the CP He-strong stars. However other very specific peculiar stars have been found magnetic: the Of?p stars, the UV variable stars, and the $\tau$~Sco-type stars showing also very peculiar UV spectral lines \citep{petitv11}. Another interesting result of the MiMeS project is the intensive investigation of the magnetosphere of the magnetic MiMeS targets which allowed to confront the Rigidly-Rotating magnetosphere of \citet{townsend05}. This model predicts the presence of centrifugal or dynamical magnetospheres around magnetic OB stars, depending on their magnetic strength and rotation period. We find that these predictions are well reproduced with the observations.

However some MiMeS results led to new questions. First, \citet{kochukhov11} have found that the magnetic field of the He-strong star HD 37776 is unusually complex compared to all other magnetic OB stars studied up to now. This result challenges the fossil field theory which is usually adopted to explain the origin of the magnetic fields in massive stars. Then, recent observations of the pre-MS Herbig Ae star HD 190073 have shown an unexpected field behaviour: while the field staid stable for 5 years, strong variations started to occur in 2011. This abrupt change is also challenging the fossil field theory (Alecian et al. in prep.). One explanation could be the fact that the star is very young, still evolving along the PMS, and the magnetic field has not yet stabilised and reached its final configuration. The work of \citet{duez11} predicts a period of stabilisation of the magnetic geometry before reaching its final and stable configuration. The magnetic field of HD 190073 may have not yet reached its definite configuration. Finally, a small fraction only of massive stars host strong magnetic fields, which is also challenging the fossil field theory.

To go further and bring new elements in the understanding of these puzzling results, three knew projects have or will start very soon. The first one is the magnetic investigation of the progenitors of the Herbig Ae/Be stars, the intermediate-mass T~Tauri stars (IMTTS), leaded by E. Alecian and G. Hussain. The IMTTS are intermediate-mass stars that evolve at the very beginning of the PMS phase, and can be either totally convective, partly convective, or had just become totally radiative. This project will help to better understand the role of the convective phase that stars experience before reaching the birthline on the fossil field. The second project is BinaMIcS (Binarity and Magnetic Interaction of various classes of Stars): a new ESPaDOnS Large Programme that has been submitted to the CFHT by E. Alecian and G. Wade in order to investigate the magnetic fields, magnetic and tidal interaction, and magnetosphere interactions inside binary systems of all masses. If this programme is accepted it will allow us, among many other aspects, to better constrain the fossil theory by establishing the incidence of magnetic fields among binary systems, and evaluate the role of the binarity on the magnetic fields of all kind of stars. Finally a new instrument is in preparation: Spirou that will be installed at the CHFT in about 5 years. Spirou is a spectropolarimeter in the IR. On a longer time scale, we intend to investigate the magnetic fields of the Massive Young Stellar Objects, the progenitors of the PMS and MS intermediate and massive stars.

\acknowledgements I wish to thank the Programme National de Physique Stellaire (PNPS) for their support.

\bibliography{invited_alecian}

\begin{thebibliography}{}
\expandafter\ifx\csname natexlab\endcsname\relax\def\natexlab#1{#1}\fi
\expandafter\ifx\csname url\endcsname\relax
  \def\url#1{\texttt{#1}}\fi
\expandafter\ifx\csname urlprefix\endcsname\relax\def\urlprefix{URL }\fi
\providecommand{\eprint}[2][]{\url{#2}}

\bibitem[{{Alecian} et~al.(2011){Alecian}, {Kochukhov}, {Neiner}, {Wade}, {de
  Batz}, {Henrichs}, {Grunhut}, {Bouret}, {Briquet}, {Gagne}, {Naze}, {Oksala},
  {Rivinius}, {Townsend}, {Walborn}, {Weiss}, \& {Mimes
  Collaboration}}]{alecian11}
{Alecian}, E., {Kochukhov}, O., {Neiner}, C., {Wade}, G.~A., {de Batz}, B.,
  {Henrichs}, H., {Grunhut}, J.~H., {Bouret}, J.-C., {Briquet}, M., {Gagne},
  M., {Naze}, Y., {Oksala}, M.~E., {Rivinius}, T., {Townsend}, R.~H.~D.,
  {Walborn}, N.~R., {Weiss}, W., \& {Mimes Collaboration} 2011, \aap, 536, L6

\bibitem[{{Alecian} et~al.(2009){Alecian}, {Wade}, {Catala}, {Bagnulo},
  {B{\"o}hm}, {Bouret}, {Donati}, {Folsom}, {Grunhut}, \&
  {Landstreet}}]{alecian09}
{Alecian}, E., {Wade}, G.~A., {Catala}, C., {Bagnulo}, S., {B{\"o}hm}, T.,
  {Bouret}, J., {Donati}, J., {Folsom}, C.~P., {Grunhut}, J., \& {Landstreet},
  J.~D. 2009, \mnras, 1250

\bibitem[{{Alecian} et~al.(2012){Alecian}, {Wade}, {Catala}, {Grunhut},
  {Boehm}, {Bouret}, {Flood}, {Folsom}, , {Landstreet}, {Marsden}, \&
  {Waite}}]{alecian12}
{Alecian}, E., {Wade}, G.~A., {Catala}, C., {Grunhut}, J.~D., {Boehm}, T.,
  {Bouret}, J., {Flood}, J., {Folsom}, C., , {Landstreet}, J.~D., {Marsden},
  S.~C., \& {Waite}, I.~A. 2012, \mnras, submitted

\bibitem[{{Babcock}(1947)}]{babcock47}
{Babcock}, H.~W. 1947, \apj, 105, 105

\bibitem[{{Bagnulo} et~al.(2012){Bagnulo}, {Landstreet}, {Fossati}, \&
  {Kochukhov}}]{bagnulo12}
{Bagnulo}, S., {Landstreet}, J.~D., {Fossati}, L., \& {Kochukhov}, O. 2012,
  \aap, 538, A129

\bibitem[{{Borra} \& {Landstreet}(1980)}]{borra80}
{Borra}, E.~F., \& {Landstreet}, J.~D. 1980, \apjs, 42, 421

\bibitem[{{Braithwaite} \& {Nordlund}(2006)}]{braithwaite06}
{Braithwaite}, J., \& {Nordlund}, {\AA}. 2006, \aap, 450, 1077

\bibitem[{{Donati} et~al.(2006){Donati}, {Howarth}, {Bouret}, {Petit},
  {Catala}, \& {Landstreet}}]{donati06}
{Donati}, J.-F., {Howarth}, I.~D., {Bouret}, J.-C., {Petit}, P., {Catala}, C.,
  \& {Landstreet}, J. 2006, \mnras, 365, L6

\bibitem[{{Donati} \& {Landstreet}(2009)}]{donati09}
{Donati}, J.-F., \& {Landstreet}, J.~D. 2009, \araa, 47, 333

\bibitem[{{Donati} et~al.(1997){Donati}, {Semel}, {Carter}, {Rees}, \& {Collier
  Cameron}}]{donati97}
{Donati}, J.-F., {Semel}, M., {Carter}, B.~D., {Rees}, D.~E., \& {Collier
  Cameron}, A. 1997, \mnras, 291, 658

\bibitem[{{Duez}(2011)}]{duez11}
{Duez}, V. 2011, Astronomische Nachrichten, 332, 983

\bibitem[{{Folsom} et~al.(2008){Folsom}, {Wade}, {Kochukhov}, {Alecian},
  {Catala}, {Bagnulo}, {B{\"o}hm}, {Bouret}, {Donati}, {Grunhut}, {Hanes}, \&
  {Landstreet}}]{folsom08}
{Folsom}, C.~P., {Wade}, G.~A., {Kochukhov}, O., {Alecian}, E., {Catala}, C.,
  {Bagnulo}, S., {B{\"o}hm}, T., {Bouret}, J., {Donati}, J., {Grunhut}, J.,
  {Hanes}, D.~A., \& {Landstreet}, J.~D. 2008, \mnras, 391, 901

\bibitem[{{Grunhut} et~al.(2009){Grunhut}, {Wade}, {Marcolino}, {Petit},
  {Henrichs}, {Cohen}, {Alecian}, {Bohlender}, {Bouret}, {Kochukhov}, {Neiner},
  {St-Louis}, \& {Townsend}}]{grunhut09}
{Grunhut}, J.~H., {Wade}, G.~A., {Marcolino}, W.~L.~F., {Petit}, V.,
  {Henrichs}, H.~F., {Cohen}, D.~H., {Alecian}, E., {Bohlender}, D., {Bouret},
  J.-C., {Kochukhov}, O., {Neiner}, C., {St-Louis}, N., \& {Townsend}, R.~H.~D.
  2009, \mnras, 400, L94

\bibitem[{{Henrichs} et~al.(2003){Henrichs}, {Neiner}, \& {Geers}}]{henrichs03}
{Henrichs}, H.~F., {Neiner}, C., \& {Geers}, V.~C. 2003, in A Massive Star
  Odyssey: From Main Sequence to Supernova, edited by K.~{van der Hucht},
  A.~{Herrero}, \& C.~{Esteban}, vol. 212 of IAU Symposium, 202

\bibitem[{{Kochukhov} et~al.(2011){Kochukhov}, {Lundin}, {Romanyuk}, \&
  {Kudryavtsev}}]{kochukhov11}
{Kochukhov}, O., {Lundin}, A., {Romanyuk}, I., \& {Kudryavtsev}, D. 2011, \apj,
  726, 24

\bibitem[{{Kochukhov} \& {Wade}(2010)}]{kochukhov10}
{Kochukhov}, O., \& {Wade}, G.~A. 2010, \aap, 513, A13

\bibitem[{{Landi degl'Innocenti}(1992)}]{landi92}
{Landi degl'Innocenti}, E. 1992, {Magnetic field measurements} (Solar
  Observations: Techniques and Interpretation), 71

\bibitem[{{Landstreet}(1970)}]{landstreet70}
{Landstreet}, J.~D. 1970, \apj, 159, 1001

\bibitem[{{Landstreet} \& {Borra}(1978)}]{landstreet78}
{Landstreet}, J.~D., \& {Borra}, E.~F. 1978, \apjl, 224, L5

\bibitem[{{Martins} et~al.(2010){Martins}, {Donati}, {Marcolino}, {Bouret},
  {Wade}, {Escolano}, {Howarth}, \& {Mimes Collaboration}}]{martins10}
{Martins}, F., {Donati}, J.-F., {Marcolino}, W.~L.~F., {Bouret}, J.-C., {Wade},
  G.~A., {Escolano}, C., {Howarth}, I.~D., \& {Mimes Collaboration} 2010,
  \mnras, 407, 1423

\bibitem[{{Moss}(2001)}]{moss01}
{Moss}, D. 2001, in ASP Conf. Ser. 248: Magnetic Fields Across the
  Hertzsprung-Russell Diagram, edited by G.~{Mathys}, S.~K. {Solanki}, \& D.~T.
  {Wickramasinghe}, 305

\bibitem[{{Naz{\'e}} et~al.(2010){Naz{\'e}}, {Ud-Doula}, {Spano}, {Rauw}, {De
  Becker}, \& {Walborn}}]{naze10}
{Naz{\'e}}, Y., {Ud-Doula}, A., {Spano}, M., {Rauw}, G., {De Becker}, M., \&
  {Walborn}, N.~R. 2010, \aap, 520, A59

\bibitem[{{Naz{\'e}} et~al.(2008){Naz{\'e}}, {Walborn}, \& {Martins}}]{naze08}
{Naz{\'e}}, Y., {Walborn}, N.~R., \& {Martins}, F. 2008, Rev. Mexicana Astron.
  Astrofis., 44, 331

\bibitem[{{Neiner} et~al.(2012){Neiner}, {Alecian}, {Briquet}, {Floquet},
  {Fr{\'e}mat}, {Martayan}, {Thizy}, \& {Mimes Collaboration}}]{neiner12}
{Neiner}, C., {Alecian}, E., {Briquet}, M., {Floquet}, M., {Fr{\'e}mat}, Y.,
  {Martayan}, C., {Thizy}, O., \& {Mimes Collaboration} 2012, \aap, 537, A148

\bibitem[{{Neiner} et~al.(2003){Neiner}, {Henrichs}, {Floquet}, {Fr{\'e}mat},
  {Preuss}, {Hubert}, {Geers}, {Tijani}, {Nichols}, \& {Jankov}}]{neiner03}
{Neiner}, C., {Henrichs}, H.~F., {Floquet}, M., {Fr{\'e}mat}, Y., {Preuss}, O.,
  {Hubert}, A.-M., {Geers}, V.~C., {Tijani}, A.~H., {Nichols}, J.~S., \&
  {Jankov}, S. 2003, \aap, 411, 565

\bibitem[{{Oksala} et~al.(2012){Oksala}, {Wade}, {Townsend}, {Owocki},
  {Kochukhov}, {Neiner}, {Alecian}, \& {Grunhut}}]{oksala12}
{Oksala}, M.~E., {Wade}, G.~A., {Townsend}, R.~H.~D., {Owocki}, S.~P.,
  {Kochukhov}, O., {Neiner}, C., {Alecian}, E., \& {Grunhut}, J. 2012, \mnras,
  419, 959

\bibitem[{{Petit} et~al.(2011{\natexlab{a}}){Petit}, {Ligni{\`e}res},
  {Auri{\`e}re}, {Wade}, {Alina}, {Ballot}, {B{\"o}hm}, {Jouve}, {Oza},
  {Paletou}, \& {Th{\'e}ado}}]{petitp11}
{Petit}, P., {Ligni{\`e}res}, F., {Auri{\`e}re}, M., {Wade}, G.~A., {Alina},
  D., {Ballot}, J., {B{\"o}hm}, T., {Jouve}, L., {Oza}, A., {Paletou}, F., \&
  {Th{\'e}ado}, S. 2011{\natexlab{a}}, \aap, 532, L13

\bibitem[{{Petit} et~al.(2011{\natexlab{b}}){Petit}, {Massa}, {Marcolino},
  {Wade}, {Ignace}, \& {Mimes Collaboration}}]{petitv11}
{Petit}, V., {Massa}, D.~L., {Marcolino}, W.~L.~F., {Wade}, G.~A., {Ignace},
  R., \& {Mimes Collaboration} 2011{\natexlab{b}}, \mnras, 412, L45

\bibitem[{{Townsend} \& {Owocki}(2005)}]{townsend05}
{Townsend}, R.~H.~D., \& {Owocki}, S.~P. 2005, \mnras, 357, 251

\bibitem[{{Townsend} et~al.(2005){Townsend}, {Owocki}, \&
  {Groote}}]{townsend05b}
{Townsend}, R.~H.~D., {Owocki}, S.~P., \& {Groote}, D. 2005, \apjl, 630, L81

\bibitem[{{ud-Doula} \& {Owocki}(2002)}]{uddoula02}
{ud-Doula}, A., \& {Owocki}, S.~P. 2002, \apj, 576, 413

\bibitem[{{Ud-Doula} et~al.(2008){Ud-Doula}, {Owocki}, \&
  {Townsend}}]{uddoula08}
{Ud-Doula}, A., {Owocki}, S.~P., \& {Townsend}, R.~H.~D. 2008, \mnras, 385, 97

\bibitem[{{Wade} et~al.(2012){Wade}, {Grunhut}, {Gr{\"a}fener}, {Howarth},
  {Martins}, {Petit}, {Vink}, {Bagnulo}, {Folsom}, {Naz{\'e}}, {Walborn},
  {Townsend}, \& {Evans}}]{wade12}
{Wade}, G.~A., {Grunhut}, J., {Gr{\"a}fener}, G., {Howarth}, I.~D., {Martins},
  F., {Petit}, V., {Vink}, J.~S., {Bagnulo}, S., {Folsom}, C.~P., {Naz{\'e}},
  Y., {Walborn}, N.~R., {Townsend}, R.~H.~D., \& {Evans}, C.~J. 2012, \mnras,
  419, 2459

\end{thebibliography}







\end{document}